\documentclass[preprintnumbers,amsmath,amssymb]{revtex4}
\usepackage{graphics,graphicx}
\usepackage{dcolumn}
\usepackage{subfigure}
\makeatletter
\begin{document}

\title{The Morphology of Urban Agglomerations for Developing Countries:
A Case Study with China}
%\titlerunning{The Morphology of Urban Agglomerations}
\author{Kausik Gangopadhyay}
\email{kausik@iimk.ac.in} \affiliation{Indian Institute of
Management Kozhikode, IIMK Campus P.O., Kozhikode 673570, India}
\author{B. Basu}
\email{banasri@isical.ac.in, Fax:91+(033)2577-3026}
\affiliation{Physics and Applied Mathematics Unit\\
 Indian Statistical Institute\\
 Kolkata-700108, India }
%
%
%\maketitle
\begin{abstract}
\begin{center}
{\bf Abstract}
\end{center}
In this article, the relationship between two well-accepted empirical propositions regarding the distribution of population in cities, namely, Gibrat's law and Zipf's law,  are rigorously examined  using the Chinese census data.
Our findings are quite in contrast with the most of the previous studies performed exclusively for developed countries. This motivates us to build a general environment to explain the morphology of urban agglomerations both in developed and developing countries. A dynamic process of job creation generates a particular distribution for the urban agglomerations and introduction of Special Economic Zones (SEZ) in this abstract environment shows that the empirical observations are in good agreement with the proposed model.
\end{abstract}

\maketitle

\section{Introduction}

Social phenomenon is a pertinent topic of discussion among the
Economists and Econophysicists - partly because, human behavior can
be explained in terms of Economic motives as well as a manifestation
of a complex natural system. One of the interesting observation is
distribution of dwellers in different urban agglomerations. A simple
empirical law,  namely Zipf's law \cite{zipf}, is often successful
in describing the distribution of populations for various
cities \cite{f1} in
a nation.
%Quantitatively speaking, it postulates that
%  the population-wise rank, $R_x$, of a city with $x$ number of inhabitants
%  is proportional to $x^{-\alpha}$ with $\alpha$ being one.
%  Empirically, this law has been found to be quite robust with few
%  exceptions.

In Economics, there is a body of literature devoted to explain
morphology of cities. The survey paper by Gabaix and Ioannides \cite{gabaix_survey}
enlists most of them. Krugman \cite{krugman96} have looked at
the top 135 U.S. cities and have found that the log-rank of a city bears a linear relation
to the log-size of the same in a significant way. The slope of the linear relation
is also found to be quite close to one as expected from the Zipf's law.

Gabaix \cite{gabaix1999a} investigate into the growth of cities
and their adherence to the Zipf's law. This is because
Zipf's law is not a static phenomenon, but is the outcome of a dynamic process.
 Different cities have presumably different growth processes. We can express
the expected growth rate of a city with population $S$ as a random variable,
$\mu(S)$. The standard deviation in the growth rate of cities with population
$S$ are denoted by $\sigma(S)$.  If either $\mu(S)$ or $\sigma(S)$ is a
non-trivial function of $S$ at least  in the upper tail of the distribution of
$S$, there would be  violations  of Zipf's law. This is a consequence of
the Gibrat's law being followed in the upper tail of the city distribution.
Gibrat's law proposes that the growth rate process of a city is independent
of the size of the city. Therefore the mean growth rate and the standard deviation of
the growth rate for a city is independent of its size. It must be clarified that
Gibrat law does not say that the growth rate of any city follows the same
stochastic process. It only says that there is no relation between growth rate
of a city and its size.

Gibrat law and its relation to Zipf's law is particularly pertinent for a nation
experiencing growth in urban inhabitants. A developing country is very different
compared to its developed counterparts in terms of economic and social structures.
Therefore, the inter relationship between
this two empirical conjectures might be particularly interesting.
A pertinent case study is the People's Republic of
China, where urbanization is taking place in a fast pace.
We investigate into the occurrence of these laws in case of China.
The next section discusses our empirical analysis with the
findings. A model is proposed in Section \ref{section:model} along
with an appropriate simulation study. The concluding remarks are noted
in Section \ref{section:discuss}.

 \section{Data Treatment}
 \label{data_treatment}

\begin{figure}[h]
\centering
\subfigure[Census year 1990: rank of a city plotted against its size]{
\includegraphics[scale=0.2]{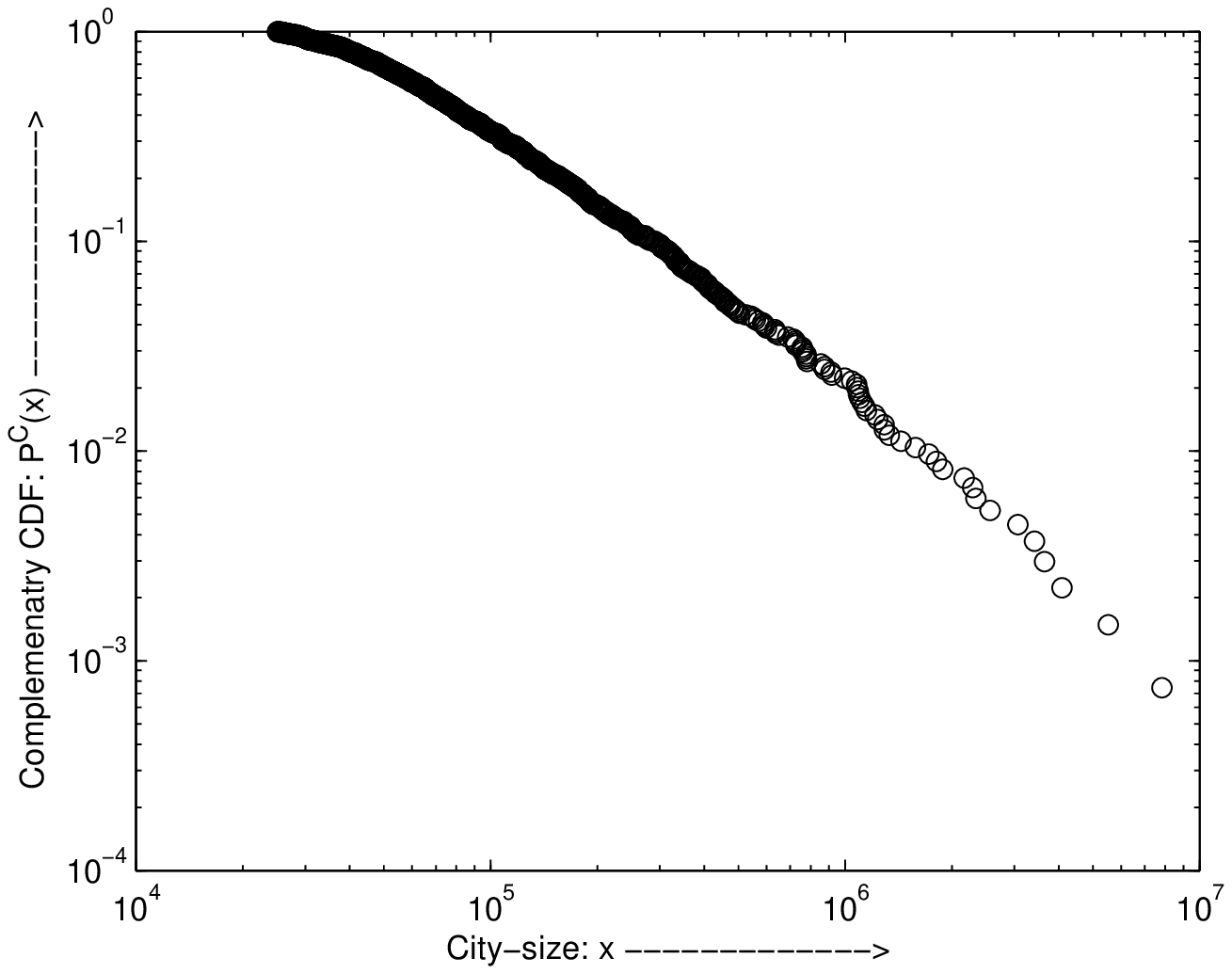}
\label{fig:census_1990}
}
\subfigure[Census year 2000: rank of a city plotted against its size]{
\includegraphics[scale=0.2]{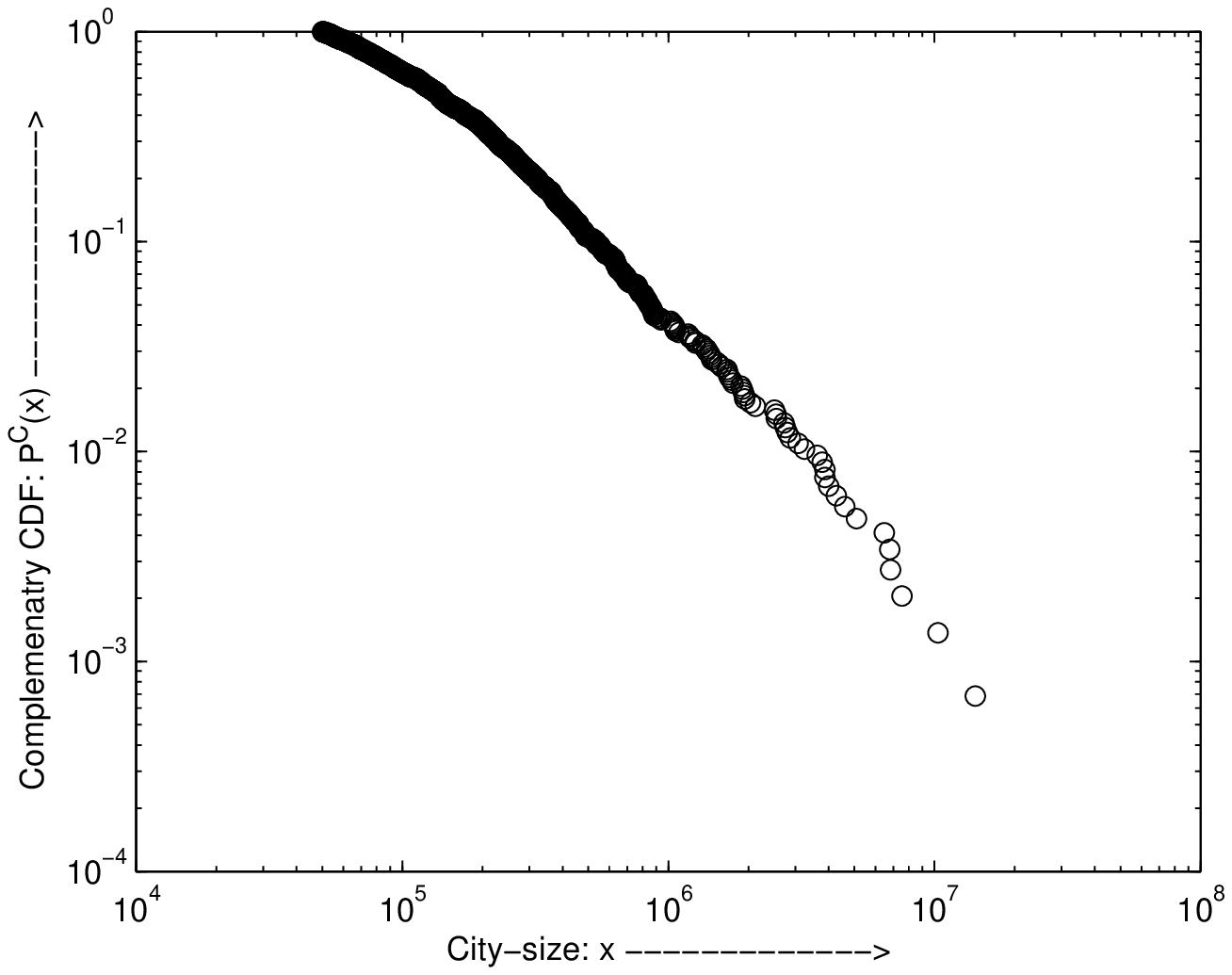}
\label{fig:census_2000}
}
\subfigure[Scatter plot of city growth against city size (1990-2000)]{
\includegraphics[scale=0.2]{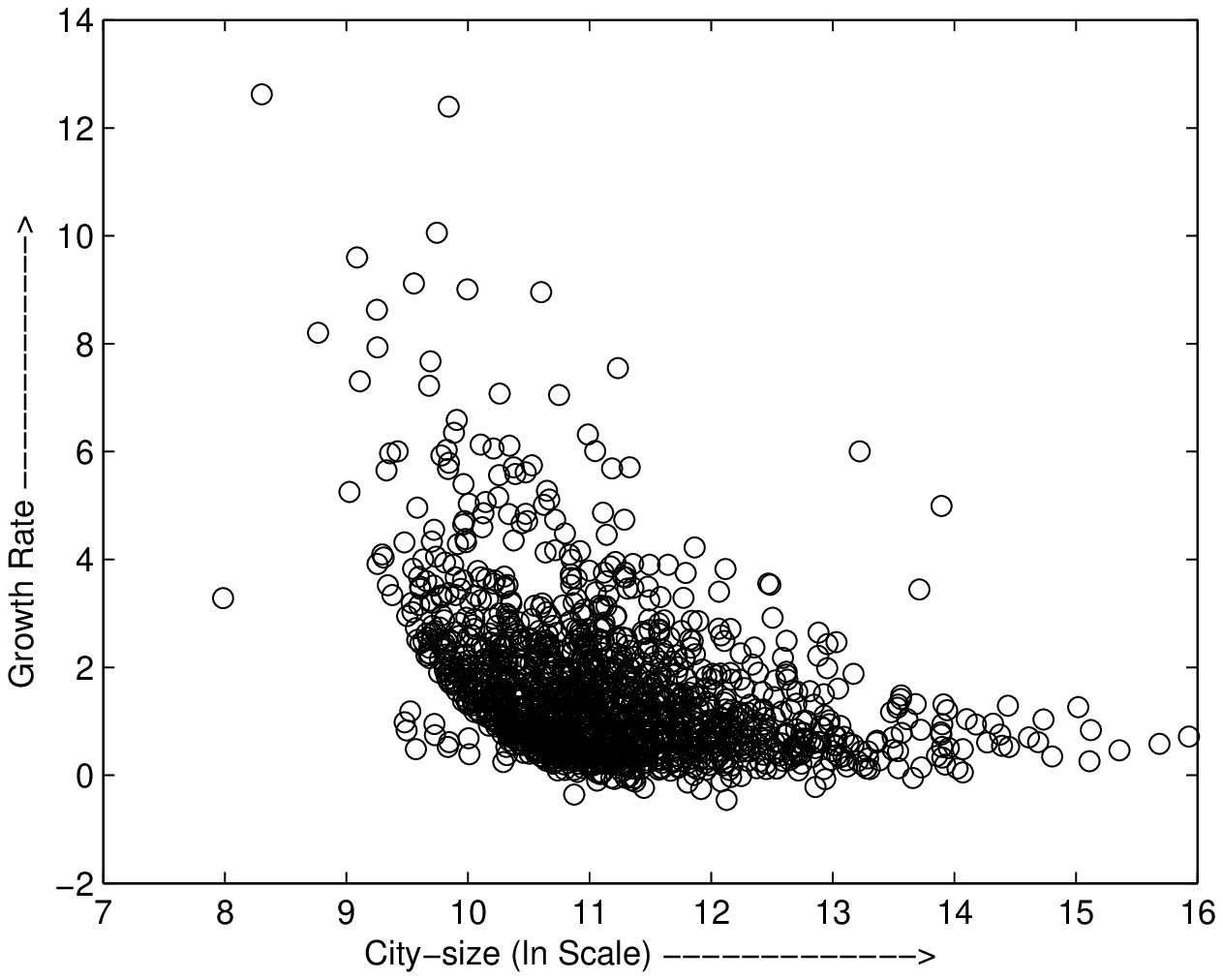}
\label{fig:scatter_plot}
}
\caption[Optional
caption for list of figures]{Chinese Cities: 1990-2000}
 \label{fig:data}
\end{figure}

 The People's Republic of China conducted censuses in 1953, 1964, and 1982.
 At the 2000 census, the total population
  stood at approximately  1.29533 billion, which is about 22\% of total
  population in the world. 36\% of the Chinese population used to reside
  in urban agglomerations in 2000. We use the data \cite{china_data} from
  1990 and 2000 census (plotted in Fig. \ref{fig:data}).

%\begin{figure}[t]
%\sidecaption[t]
%\includegraphics[scale=.3]{gibrat_china_scatter.eps}
%
% If no graphics program available, insert a blank space i.e. use
%\picplace{5cm}{2cm} % Give the correct figure height and width in cm
%
%\caption{Scatter plot of city growth against city size}
%\caption{}
%\label{fig:Scatter}       % Give a unique label
%\end{figure}

 \subsection{Verification of Zipf's Law}
 %The notations are described here for clarity of our methods.
Let $p(\cdot)$ be a probability density function of the city-size distribution.
 The corresponding cumulative distribution
  function (CDF) and the complementary cumulative distribution function (CCDF)
  are given by $P(\cdot)$ and $P^C(\cdot)$,
  respectively. %The CDF, $P(x)$ is the probability that a city has a population less than or equal to x and the CCDF, $P^C(x)$, is the  probability that a
%city has a population greater  than x.
  By definition,
\begin{equation*}
 P(x)=\int_0^x
p(x^\prime) dx^{\prime};~~~~P^C(x) = 1 - P(x)
\end{equation*}

In  case of city-size distribution following the Zipf's law,
\begin{equation}
p_{\alpha}(x) = Cx^{-\alpha}~~~ \mbox{and}~~~ \displaystyle
P_{\alpha}^C(x)=\frac{C}{\alpha-1} x^{-(\alpha-1)} \label{Pareto}
\end{equation}
where $\alpha$ and $C$ are constants. $\alpha$ is called the
exponent of the power law. This family of power law distributions
for $\alpha
>1$ are known as the Pareto distribution. From equation (\ref{Pareto}),
it is obvious that $p_{\alpha}(x)$ diverges to infinity for any
value of $\alpha
> 1$ as $x \rightarrow 0$. Therefore, some minimum value, $x_{min}$,
is usually considered for the support of the Pareto distribution.
%The corresponding probability density function, the CDF and the CCDF
%are given by:
%\begin{equation}
%p_{\alpha}(x) = \frac{(\alpha-1)\cdot
%x_{min}^{\alpha-1}}{x^{\alpha}};~~~~~~~P_{\alpha}(x) = 1 -
%\left(\frac{x_{min}}{x}\right)^{\alpha-1};~~~~~~~P_{\alpha}^C(x) =
%\left(\frac{x_{min}}{x}\right)^{\alpha-1} .\label{Pareto_dist}
%\end{equation}

\begin{table}[h]
\caption{\textbf{Data Description}: Values of city-population
are reported in units of thousands. The left truncation of the data
is determined through the value of $x_{min}$. The numbers in parenthesis
represent the standard errors for the estimates
Source: \cite{china_data}
}
\label{table:data}
\begin{tabular}{cccccccc|cc}
\hline \hline
    Census & n & Min & Max & Mean & Median &  First & Third &  \multicolumn{2}{|c}{Estimate of $\alpha$}  \\
\cline{9-10}
 Year & & Value & Value & & & Quartile  & Quartile  & Linear Fit &
MLE  \\
\hline  2000 & 1462 & 50.08 & 14230.99 & 298.27 & 136.63 &
80.86 & 265.42 & 1.7544 & 2.2975
\\
 & & & & & & & & (0.0018) & (0.0572)
\\
1990 & 1345 & 25.02 & 7821.79 & 156.33 & 68.71
& 44.23 & 128.96 & 1.7701 & 2.2308
\\
 & & & & & & & & (0.0032) & (0.0736)
\\
\hline \hline
\end{tabular}
\end{table}

The slope of the plot, in which log of the rank of a city,
$\log(R_x)$, is plotted against the log of its population,
$\log(x)$, has been used to estimate the exponent of the power law
in almost all the previous studies. It has been shown
\cite{shalizi_powerlaw, gabaix1999a} that this produces a biased estimate of the
power law exponent. Alternatively the Maximum Likelihood Estimator\cite{f2}{The MLE is given by the expression, $\widehat{\alpha}_{MLE} = 1+ n \left[\sum_{i=1}^n
\log\left(\frac{x_i}{x_{min}}\right)\right]^{-1}$}
(MLE) produces the most efficient estimate. We find \cite{physica_2009} the estimate of $\alpha$ to be significantly bigger than
2 as a departure from the Zipf's law (see Table \ref{table:data}).

\subsection{Verification of Gibrat's Law}

The cities in the upper tail of the size distribution
follow a constant  rate of growth for various developed countries \cite{JanE_2004}. It is interesting to
repeat this exercise for a developing nation, where urbanization is happening fast
to notice any discrepancy among cities in terms of growth regarding size. We perform various non-parametric
as well as parametric exercises on the data to find out the relationship between the size of of a city and its growth rate.

\begin{figure}[h]
\centering
\subfigure[Epanechnikov Kernel]{
\includegraphics[scale=0.35]{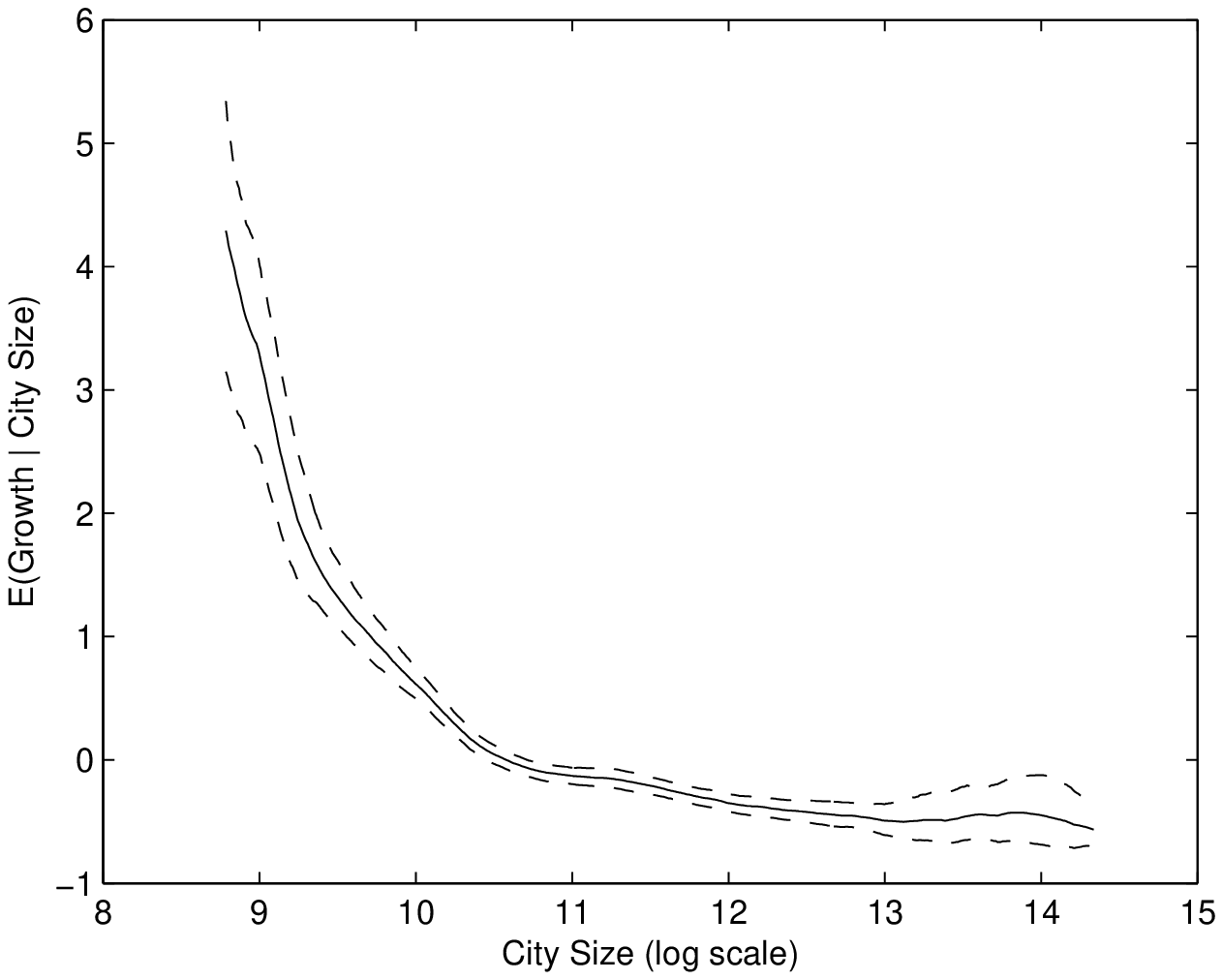}
\label{fig:kernel_mean_ep} }
\subfigure[Gaussian Kernel]{
\includegraphics[scale=0.35]{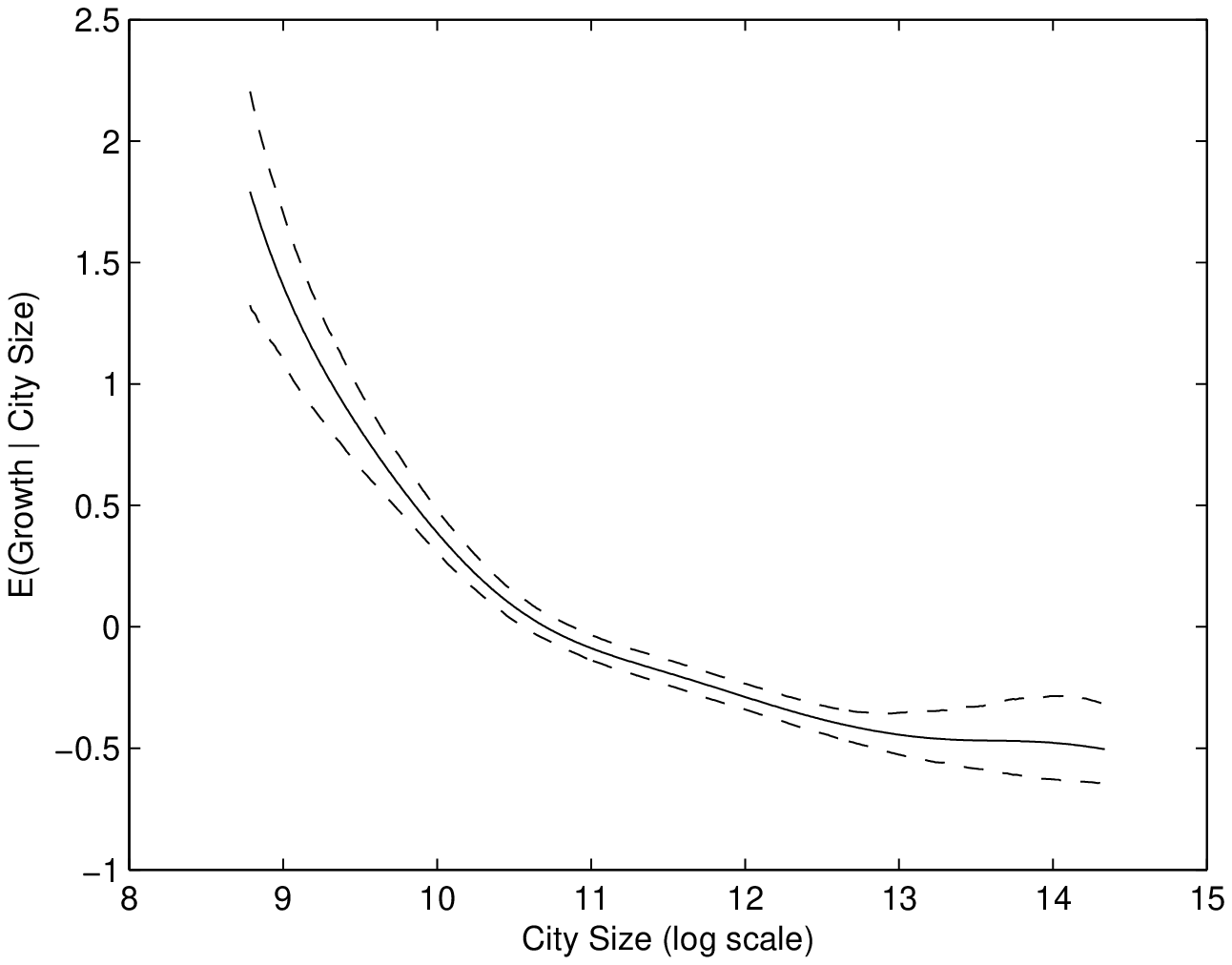}
\label{fig:kernel_mean_gauss} }
\caption[Fig 1]{Kernel estimates of population growth
against city-size
(dotted line represents the 95\% bootstrapped confidence interval)}
 \label{fig:kernel_mean}
\end{figure}

 We plot the growth rate of population in all available urban agglomerations
for the period of 1990-2000 against the population of the corresponding urban
agglomeration in 1990.  The standard non-parametric measure is to use the Kernel
estimates of local mean. Suppose, the growth rate of a city, $g_i$, bears some
relation with the size of the city, $S_i$, modeled as:

\[
g_i = m(S_i) + \epsilon_i
\]
for all $i = 1, 2, ..., n$, $n$ being the total number of cities with available data.
The objective is to find a smooth estimate of local means of growth rate over size
and to verify whether there is any visible relationship between growth and size based
on this estimate $m(\cdot)$. $g_i$ is the growth rate of the $i$th city over 1990-2000.
We perform a Kernel density regression in the support of $S_i$.\cite{f4} The local average smooths around the point $s$, and the
smoothing is done using a kernel, i.e. a continuous weight function symmetric around $s$.
The bandwidth $h$ of a kernel determines the scale of smoothing. The Nadaraya-Watson
estimate \cite{Pagan_Ullah} of $m(\cdot)$  is given by the following expression,
\[
\hat{m}(s) = \frac{n^{-1} \sum_{i=1}^{n} K_h(s - S_i) g_i}{n^{-1} \sum_{i=1}^{n} K_h(s - S_i)}
\]

We use two most popular Kernels, Gaussian and Epanechnikov. For Gaussian Kernel, $K(\psi) = (2 \pi)^{-1/2} \exp\left[-\frac{1}{2}(\psi)^{2}\right]$, and for
the Epanechnikov Kernel, $K(x) = \frac{3}{4}\left(1-\psi^2\right) \cdot \mathbf{1}_{|\psi| \leq 1}$. For both the kernels, we find that $m(\cdot)$ does depend on
the size. The visual observation is verified through the following regression, where the growth
rate of a city is regressed on the size of the city \cite{f3}. We find a significant \cite{f5} negative coefficient
for the variable of city-size.

\[
\begin{array}{l}
g_i = 2.635 ~~-~~ 4.681 \times 10^{-7} \cdot \frac{S_{90}+S_{00}}{2} \\
~~~~~(0.039)~~~~~(8.982 \times 10^{-8})
\end{array}
\]

\begin{figure}[h]
\centering
\subfigure[Epanechnikov Kernel]{
\includegraphics[scale=0.3]{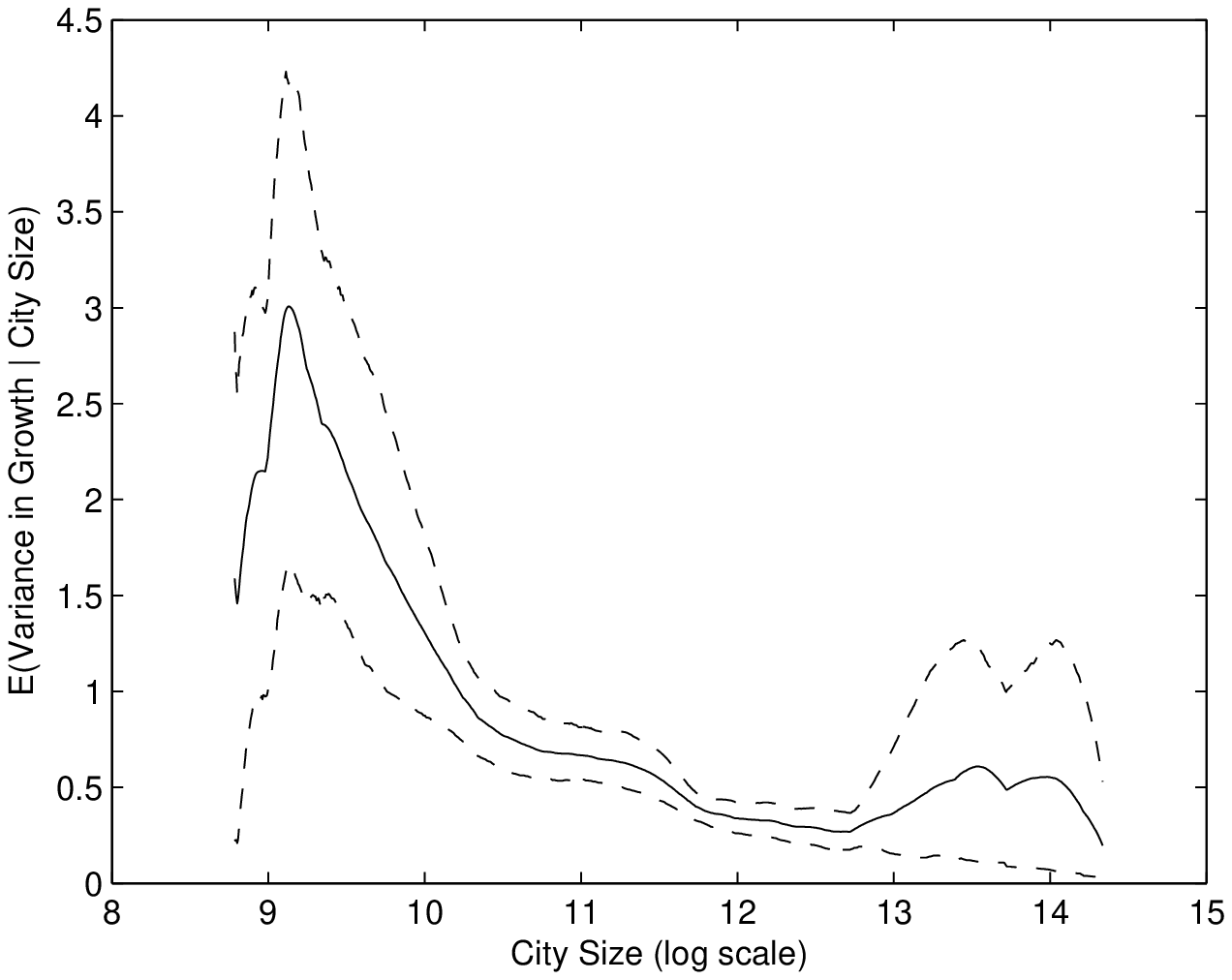}
\label{fig:kernel_var_ep} }
\subfigure[Gaussian Kernel]{
\includegraphics[scale=0.3]{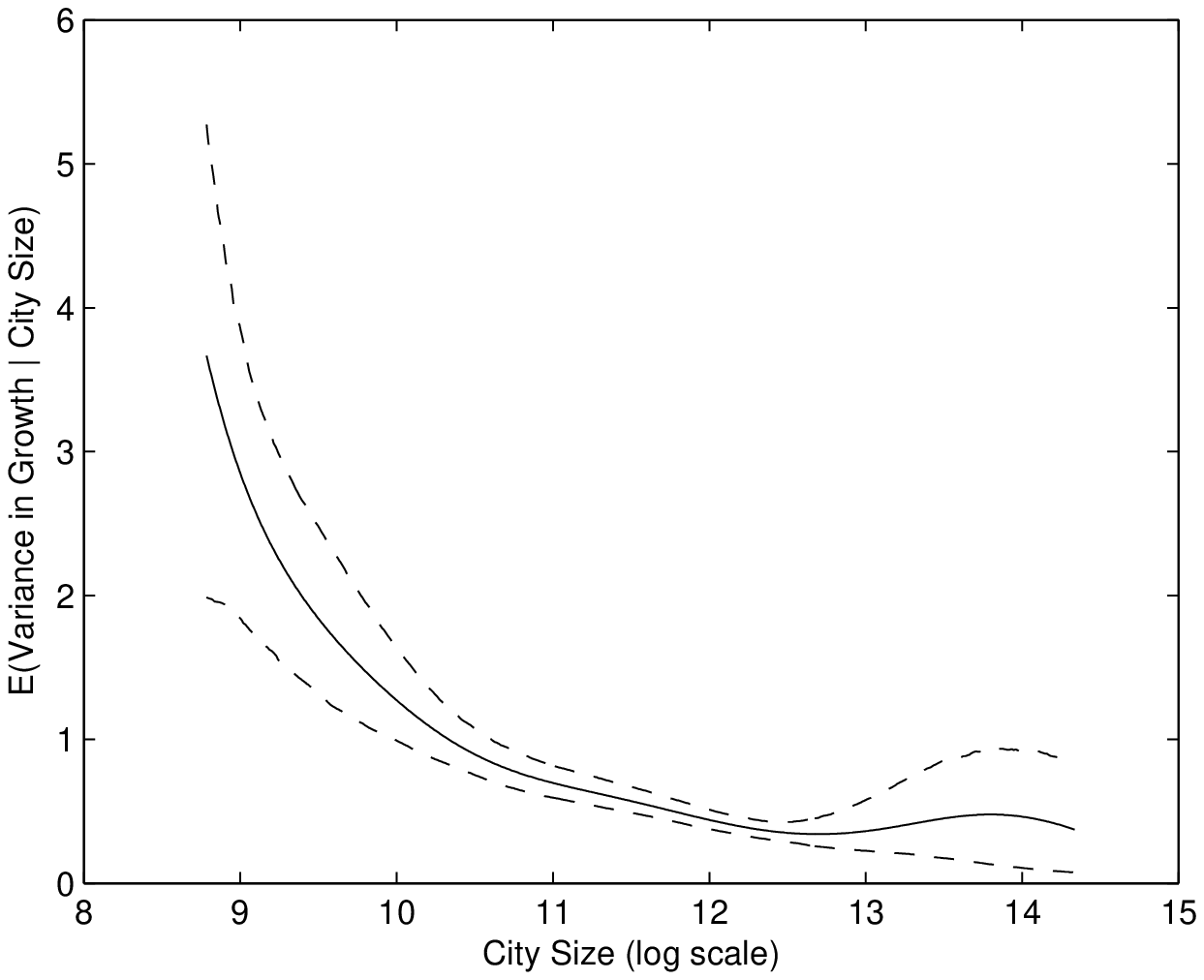}
\label{fig:kernel_var_gauss} }
\caption[Optional
caption for list of figures]{Kernel estimates of variances in
population growth against city-size
(dotted line represents the 95\% bootstrapped confidence interval)}
 \label{fig:kernel_var}
\end{figure}

We conclude that there is a definite variation among cities in terms of
growth process and the overall evidence indicates that the growth process
is negatively biased against the cities of higher sizes at least at the
upper tail of the distribution.

\section{A Migration Based Model}
\label{section:model}
To illustrate the empirical anomalies found in the context of distribution of urban agglomerations in China, we can motivate our findings with a mathematical model of city formation. There are several recent attempts \cite{batty,beng,witt} to model urban growth. It uses the idea that the growth of cities resembles to that of the two-dimensional aggregates of particles. There are results in the the statistical physics of clusters regarding the growth of the two-dimensional aggregates of particles. These results are applied in the context of modeling the population distribution of urban agglomerations. In particular, the model of diffusion limited aggregation(DLA) predicted the existence of only one large fractal cluster that is almost perfectly screened from incoming development units so that almost all the cluster growth occurs in the extreme peripheral tips. The morphology of cities is also explained using a percolation model\cite{makse}, where the scaling of the urban perimeter of individual cities and the  distribution of system of cities are tested. The intermittency mechanism\cite{yb} is used to model\cite{zan} a large scale city formation and understand the universal properties of the social phenomenon of city formation and global demographic development. In a different approach\cite{her}, the laws of population growth is explained using the \emph{City Clustering Algorithm} (CCA). The CCA is used to examine Gibrat's law of proportional growth and finds that the mean growth rate of a cluster exhibits deviations from the Gibrat's law.

For China, we need a model that is consistent with the empirical phenomenons
observed and yet models the violations of the power law  as found in the data.
However, it must be taken into account that in the developed countries, this
empirical observations are often reversed as we have found out from the literature.
We introduce the aspect of Special Economic Zones in my model and explain the empirical
anomalies in contrast to the developed countries in terms of Special Economic Zones.
We construct a baseline environment without any Special Economic Zones. Then we add Special Economic
Zones to that environment to observe any effect due to introduction of SEZ.\footnote{
A Special Economic Zone (SEZ) is a geographical region that has economic laws that
are more liberal than a country's typical economic laws. The category 'SEZ' covers
a broad range of more specific zone types, including Free Trade Zones (FTZ), Export
Processing Zones (EPZ), Free Zones (FZ), Industrial Estates (IE), Free Ports,
Urban Enterprize Zones and others. Usually the goal of a structure is to increase
foreign investment. One of the earliest and the most famous Special Economic Zones were
found by the government of the People's Republic of China under Deng Xiaoping in the
early 1980s. The most successful Special Economic Zone in China, Shenzhen, has
developed from a small village into a city with a population over ten million within 20 years.
}

There are $k$ locations in a country. Jobs are spawn one at a time. The probability of a job being spawn
in a location is a function number of already existing jobs in that location.  More particularly, the probability of an additional job being created at the $i^{th}$ location is proportional to $n_i^\gamma$, where $n_i$ is the number of already existing jobs at the $i^{th}$ location. We let jobs spawn at different location until total number of jobs becomes $N$. The parameter $\gamma$ is an important parameter of scale. If $\gamma$ is 1, the growth rate
of a city is independent of its size. On the other hand, if $\gamma$ is less than unity, larger cities are discriminated against regarding growth. A value of $\gamma$ being more than one means that
the growth process favours the large cities to growth against the smaller cities.

We introduce a migration based Special Economic Zones in this model. The government introduce the
feature of Special Economic Zones by giving special privileges to some cities. The privileged urban agglomerations
are chosen in such a way that they are not from the most populous cities. A number of new jobs are created in the
locations of the SEZs. These new jobs require higher skill levels compared to the previously existing jobs. A worker matched with these jobs leave their old locations of work and move to the new location. Also higher skilled workers are primarily from the top ranking cities.

\subsection{A Simulation Study}
To evaluate the performance of our economically tenable model, we resort to the widely used technique of simulation.
We choose 3,000 locations ($k$) and one million agents ($N$). Jobs are spawn randomly in various locations are defined in our framework until the total number of spawned jobs is equal to total number of agents. We choose the value of $\gamma$ to be 0.9 so that there is a negative bias towards the growth of top ranking cities as observed as observed in the data. We consider the top 2,500 locations and estimate the power law coefficient using the maximum likelihood method.  we find $\hat{\alpha}_{MLE}$ to be 1.0419 with standard error of the estimate being 0.0208.This baseline study is devoid of any SEZ and is quite in accordance with the Zipf's Law.

\begin{figure}[h]
\centering
\subfigure[\emph{Before} introduction of SEZs,
city sizes plotted against corresponding ranks]{
\includegraphics[scale=0.3]{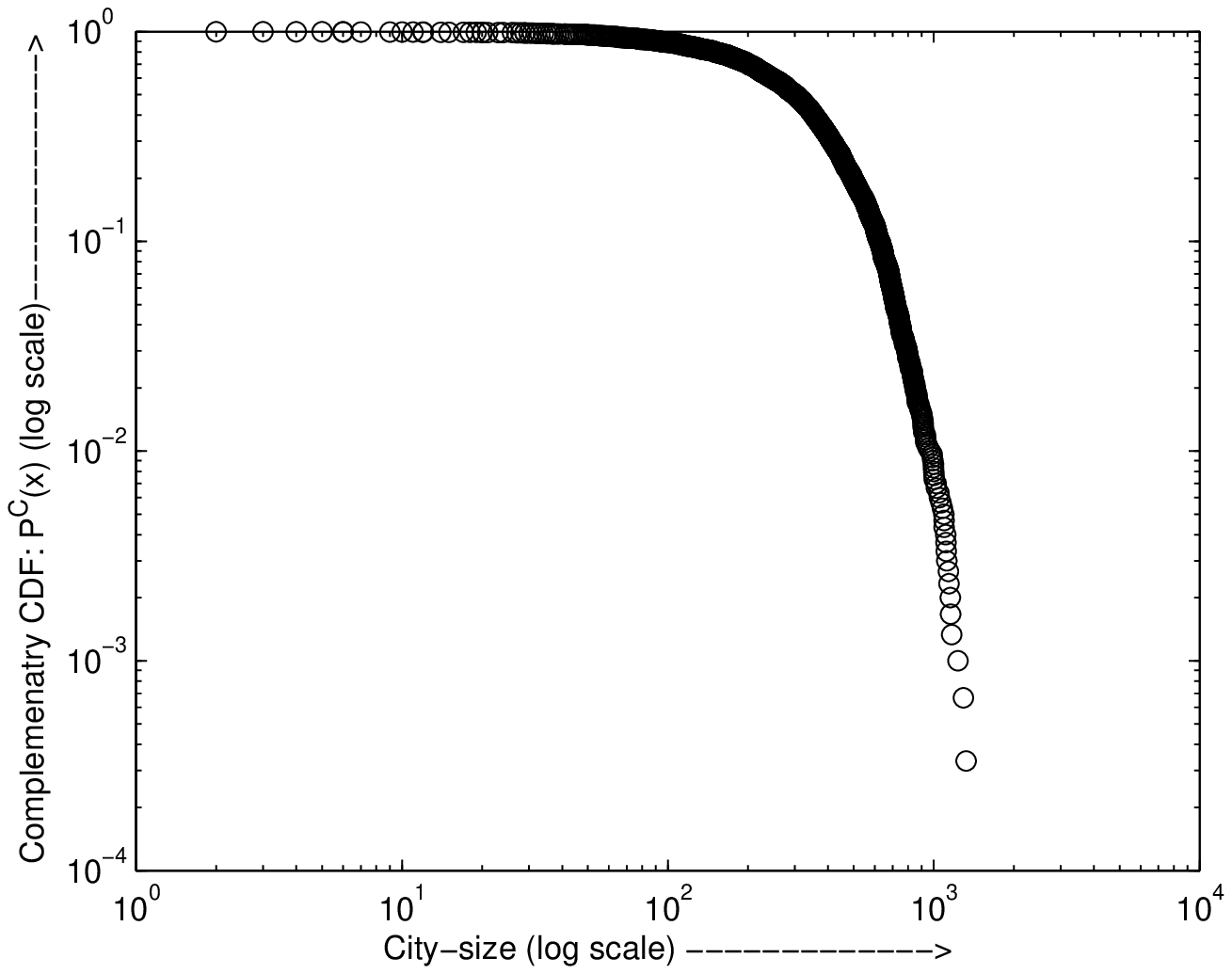}
\label{fig:before_sez} }
\subfigure[\emph{After} introduction of SEZs,
city sizes plotted against corresponding ranks]{
\includegraphics[scale=0.3]{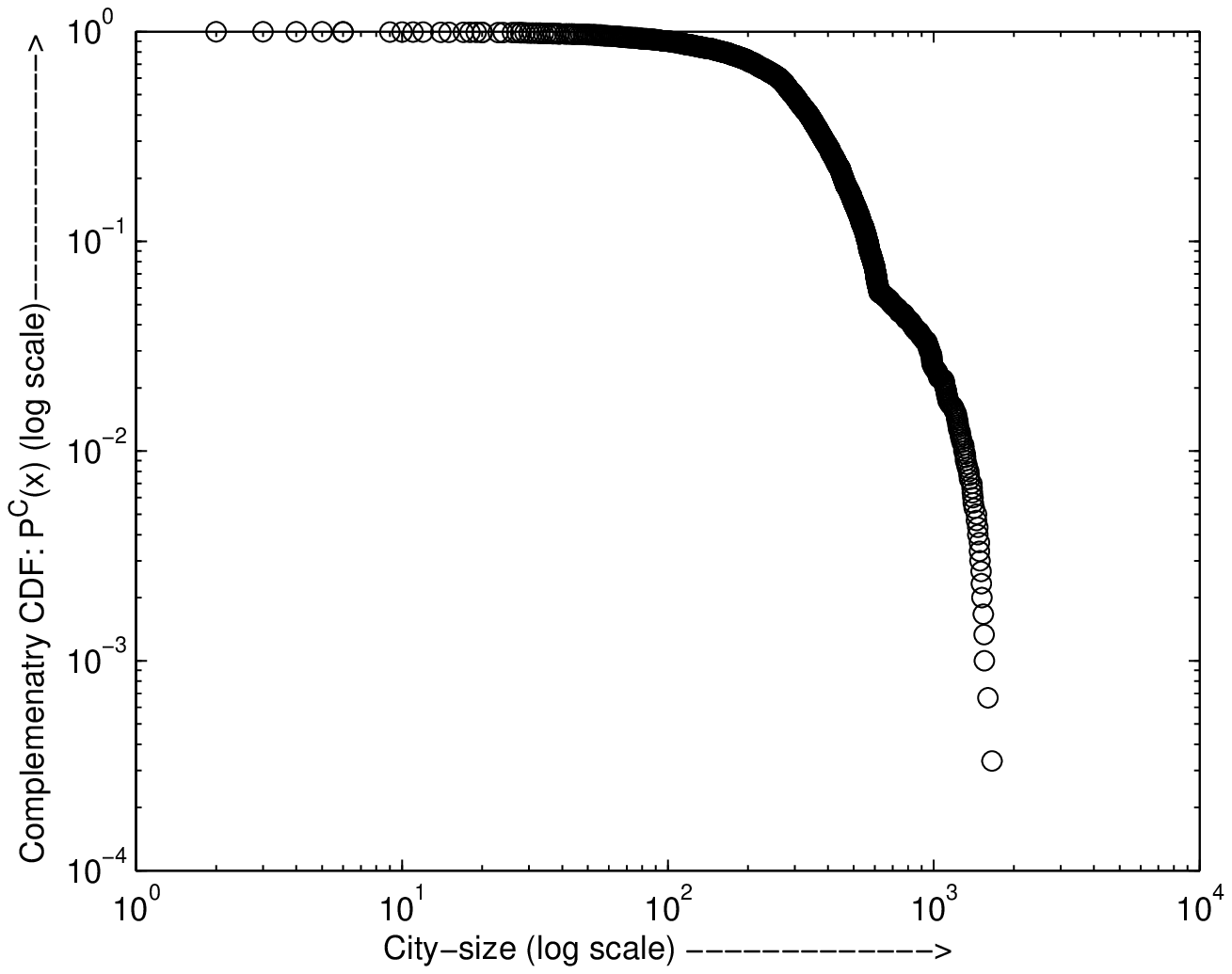}
\label{fig:after_sez} }
 \label{fig:model}
\caption[Optional
caption for list of figures]{Simulation study for the model}
\end{figure}

To introduce SEZ in this model, we randomly select 270 locations outside the top 300 locations and introduce a
number of new jobs in those locations equaling 20\% of already existing jobs in the economy.\footnote{There is nothing special about the numbers used in this constructed model. A numerical experiment with different
values for the parameters would qualitatively yield the same response.}
 Workers from the
top 300 locations are randomly matched with the newly created jobs and once matched, they migrate to the
location of their new jobs. We compute $\hat{\alpha}_{MLE}$  in the same way considering top ranking 2,500 locations
and find it to be 1.2667 with 0.0259 to be the standard error of the estimate. This is demonstrative of the high
value of $\alpha$ estimated using the data for China. Moreover, $\alpha$ estimated for the census year of 2000
is higher than that for the census year of 1990. It is associated with the rising importance of SEZs in the Chinese
economy.

%\newpage

\section{Discussion}
\label{section:discuss}

Economists often surmise\cite{gabaix_survey} that Zipf's law is the consequence of  Gibrat's law as far as city-size distribution is concerned. A simultaneous violation of both
  is natural. However, Gibrat's law is associated with the free
  market economy\cite{juan_carlos_cordoba}. A breech in Gibrat's
  law implies a wedge in the free market.
  A possible source of this wedge is debatable. We focus on government's intervention
  on the natural process of morphology of cities.  The cities under SEZ are subject to
  very different economic regulations compared to their counterparts in the rest of the country.
 This is analogous to a wedge in a perfectly competitive economic system.

 It has been pointed out\cite{JanE_2004} that the Zipf's exponent does depend on the cut-off in the upper tail
 of the city size distribution. The difference in socio-economic structure may give rise to different values of the Zipf's exponent with the same minimum cut-off. It is observed that in case of China, the exponent of Zipf's law
 augments for the year of 2000 compared to the value in the year of 1990. However, number of locations above the minimum cut-off are quite close (see Table \ref{table:data}). This phenomenon cannot be explained by a static process as modeled in \cite{JanE_2004}. Nevertheless, our model reconciles this empirical scenario with the gradual importance of SEZs in China.

%\end{document}

\begin{thebibliography}{*99}
\bibitem{f1} In
this article, ``Urban Agglomeration'' and ``City'' have throughout
been used interchangeably. The literature starting from the Zipf's law
have historically looked into the population distribution of the urban areas
formally denoted as ``Cities''. However the more general notion of urban
agglomerations have been used in the relatively recent literature.
\bibitem{f2} The MLE is given by the expression, $\widehat{\alpha}_{MLE} = 1+ n \left[\sum_{i=1}^n
\log\left(\frac{x_i}{x_{min}}\right)\right]^{-1}$
\bibitem{f4}We have chosen the
interval $[1.1 \cdot \underset{i}{\min}~ S_i,~~ 0.9 \cdot \underset{i}{\max}~S_i ]$ to exclude the effect of
the boundaries to some extent.
\bibitem{f5} The standard errors of the estimates are shown in the parenthesis below.
\end{thebibliography}
\end{document}